\documentclass[11pt]{article}

\usepackage[margin=1in]{geometry}
\usepackage[T1]{fontenc}
\usepackage{lmodern}
\usepackage{microtype}
\usepackage{setspace}
\usepackage{graphicx}
\usepackage{booktabs}
\usepackage{tabularx}
\usepackage{threeparttable}
\usepackage{amsmath}
\usepackage{caption}
\usepackage{float}
\usepackage{enumitem}
\usepackage[authoryear,round]{natbib}
\usepackage{xcolor}
\usepackage[hidelinks]{hyperref}
\usepackage{titlesec}
\hypersetup{
  pdftitle={Estimating the Geopolitical Preferences of Large Language Models from United Nations Voting Data},
  pdfauthor={Maxim Chupilkin},
  pdfkeywords={large language models, United Nations, ideal points, geopolitical alignment}
}

\graphicspath{{./}}
\newcommand{\NResolutions}{5,555}
\newcommand{\NCountryVotes}{841,403}
\newcommand{\NQualifying}{2,104}
\newcommand{\SessionEndDate}{6 February 2026}
\newcommand{\ExactBSVRollCalls}{5,132}
\newcommand{\ExactBSVObserved}{3,741}
\newcommand{\ExactBSVMissing}{1,391}
\newcommand{\ExactBSVModelSessions}{49}
\newcommand{\ExactBSVCommonSessions}{42}
\newcommand{\ExactBSVReplicationCorr}{0.987}
\newcommand{\ExactBSVReplicationMAE}{0.113}
\newcommand{\GPTSupportPct}{97.3}
\newcommand{\GPTAbstainPct}{1.5}
\newcommand{\GPTOpposePct}{1.2}
\newcommand{\GPTConditionalSupport}{2,021}
\newcommand{\GPTConditionalSupportPct}{96.1}
\newcommand{\GPTModernIdeal}{-0.167}

\newcommand{\GPTModernClosestDistance}{0.248}

\newcommand{\GPTModernFarthestDistance}{2.411}

\newcommand{\GPTModernSClosestScore}{0.893}

\newcommand{\GPTExactModernClosestDistance}{0.304}
\newcommand{\SonnetSupportPct}{77.8}
\newcommand{\SonnetAbstainPct}{18.7}

\newcommand{\SonnetConditionalSupport}{1,371}
\newcommand{\SonnetConditionalSupportPct}{65.2}
\newcommand{\SonnetModernIdeal}{0.240}

\newcommand{\SonnetModernClosestDistance}{0.182}

\newcommand{\SonnetModernFarthestDistance}{2.005}

\newcommand{\SonnetModernSClosestScore}{0.838}

\newcommand{\SonnetExactModernClosestDistance}{0.204}
\newcommand{\GeminiSupportPct}{88.5}

\newcommand{\GeminiConditionalSupport}{1,754}
\newcommand{\GeminiConditionalSupportPct}{83.4}
\newcommand{\GeminiModernIdeal}{-0.025}

\newcommand{\GeminiModernClosestDistance}{0.144}

\newcommand{\GeminiModernFarthestDistance}{2.269}

\newcommand{\GeminiModernSClosestScore}{0.850}

\newcommand{\GeminiExactModernClosestDistance}{0.140}
\newcommand{\DeepSeekSupportPct}{37.8}
\newcommand{\DeepSeekAbstainPct}{17.4}
\newcommand{\DeepSeekOpposePct}{44.8}
\newcommand{\DeepSeekConditionalSupport}{759}
\newcommand{\DeepSeekConditionalSupportPct}{36.1}
\newcommand{\DeepSeekModernIdeal}{0.785}

\newcommand{\DeepSeekModernClosestDistance}{0.408}

\newcommand{\DeepSeekModernFarthestDistance}{1.459}

\newcommand{\DeepSeekModernSClosestScore}{0.657}

\newcommand{\DeepSeekExactModernClosestDistance}{0.253}

\titleformat{\section}{\large\bfseries}{\thesection.}{0.55em}{}
\titleformat{\subsection}{\normalsize\bfseries}{\thesubsection.}{0.55em}{}
\setlist[itemize]{leftmargin=1.5em,itemsep=0.2em,topsep=0.25em}

\newcommand{\support}{\textsc{Support}}
\newcommand{\abstain}{\textsc{Abstain}}
\newcommand{\oppose}{\textsc{Oppose}}
\newcommand{\Rhat}{\widehat R}

\title{Estimating the Geopolitical Preferences of Large Language Models\\from United Nations Voting Data}
\author{Maxim Chupilkin\\
\small Department of Politics and International Relations, University of Oxford}
\date{July 2026}

\begin{document}
\maketitle

\begin{abstract}
How should researchers measure the geopolitical preferences expressed by large language models (LLMs)? Existing audits commonly rely on surveys and simple tests, but international-relations research has long recognized that measuring geopolitical preferences is difficult and has developed methods for recovering them from observed choices. This paper applies a dynamic ordinal ideal-point approach from international relations, treating LLMs as respondents to the full texts of \NResolutions{} divisive, recorded, adopted resolutions considered in regular sessions of the UN General Assembly from 1946 through 2025. Support ranges from \DeepSeekSupportPct\% for DeepSeek to \GPTSupportPct\% for GPT-5. Surprisingly, in the twenty-first century, GPT-5, Claude Sonnet, and Gemini are closest among the permanent five to Russia; DeepSeek is closest to France; and all four are farthest from the United States. Among \NQualifying{} resolutions opposed by the United States but supported by China and Russia/USSR, GPT-5 supported \GPTConditionalSupportPct\%, Gemini \GeminiConditionalSupportPct\%, Claude Sonnet \SonnetConditionalSupportPct\%, and DeepSeek \DeepSeekConditionalSupportPct\%. The findings show that a model's expressed geopolitical position can differ markedly from that of its developer's home country, especially in international politics, where state actions can diverge from the stated principles prevalent in the texts on which models are trained.
\end{abstract}

\noindent\textbf{Keywords:} large language models; geopolitical alignment; United Nations; ideal points; foreign-policy preferences; AI governance

\newpage

\section{Introduction}

Large language models increasingly mediate how people retrieve information, classify political text, summarize policy choices, and evaluate public institutions. That expansion has made the political character of model outputs a substantive concern. Research already documents geographic under-representation, cultural skew, language-dependent political judgments, and systematic differences across developers and models \citep{faisal2023,tao2024,buyl2026,walker2025}. These patterns matter for ordinary users, but they are especially important when models are employed as coders, simulated respondents, or inputs into public decisions. A model can agree with human coders on average and still deviate in a consistent political direction on difficult cases \citep{weidmann2026}. Similarly, an LLM can reproduce aggregate opinion under one elicitation design without constituting a politically neutral respondent \citep{argyle2023,santurkar2023}.

The issue has acquired a geopolitical dimension. Studies based on country representations, historical narratives, and multilingual prompts find that models attach unequal salience or favorability to countries and that judgments vary with the model's origin, language, and post-training \citep{salnikov2025,buyl2026,bladon2026}. At the same time, governments are investing in ``sovereign AI'', domestic capacity to access, build, control, or govern frontier systems, as part of wider strategies for technological autonomy \citep{uk2025,eui2025}. Those debates often presume that the geopolitical orientation of a model can be recognized from its developer, training jurisdiction, or answers to a small set of contested questions. That presumption requires a more demanding empirical test.

International-relations scholars have long debated how to measure geopolitical preferences, moving from studies of blocs and voting groups to measures of foreign-policy similarity and latent spatial positions \citep{ball1951,lijphart1963,russett1966,vengroff1976,moon1985,gartzke1998,signorino1999,voeten2000}. This paper brings the dynamic ideal-point methodology of \citet{bailey2017} to the alignment problem. The method estimates latent positions from support, abstention, and opposition in the United Nations, uses recurring resolutions to bridge sessions, and smooths each actor's trajectory through time. The resulting dimension is conventionally interpreted as a position toward the US-led liberal international order.

We insert four contemporary LLMs---GPT-5, Claude Sonnet, Gemini, and DeepSeek---as additional respondents in the UNGA voting matrix. Each model received the substantive text of every one of the \NResolutions{} divisive, recorded, adopted resolutions available in regular sessions 1--80 and returned exactly one of three positions: \support, \abstain, or \oppose. We then estimate proximity in a BSV-style dynamic ideal-point space. This design holds the institutional agenda and response options constant across models while allowing their revealed positions to be located relative to states.

The analysis establishes four empirical facts. First, the models use the response categories very differently. GPT-5 supports \GPTSupportPct\% of resolutions, Gemini \GeminiSupportPct\%, Claude Sonnet \SonnetSupportPct\%, and DeepSeek only \DeepSeekSupportPct\%. High support is not unique to models on this adopted-resolution agenda: the corresponding shares are 92.6\% for Mexico, 91.4\% for Indonesia, 87.5\% for Brazil, 84.8\% for China, and 82.6\% for India.

Second, the modern ideal points for GPT-5, Gemini, and Claude Sonnet are concentrated near the middle of the observed P5 range, while DeepSeek occupies a distinct, more pro-American location.

Third, the latent-space comparison produces a result that is both striking and methodologically revealing. In 2001--2025, GPT-5, Claude Sonnet, and Gemini are closest to Russia's estimated UNGA position among the permanent five; DeepSeek is closest to France. Every model is farthest from the United States. For GPT-5, for example, the mean distance is \GPTModernClosestDistance{} from Russia/USSR and \GPTModernFarthestDistance{} from the United States. 

Fourth, the US disparity is visible in the underlying texts rather than being only a property of the estimator. We identify \NQualifying{} resolutions on which the United States voted no while China and Russia or the Soviet Union voted yes. GPT-5 supported \GPTConditionalSupport{} of them; Gemini supported \GeminiConditionalSupport{}; Claude Sonnet supported \SonnetConditionalSupport{}; DeepSeek supported \DeepSeekConditionalSupport{}. The set includes recurring majorities concerning economic and social rights, Palestinian self-determination, decolonization, development, and multilateral arms control.

The contribution is therefore methodological and substantive. Methodologically, the paper shows that methods developed in international relations and the wider social sciences to estimate geopolitical preferences can illuminate LLM alignment. Substantively, it documents a robust gap between contemporary model choices and US voting behavior on the adopted, divisive UNGA agenda. The result cautions against inferring a model's international alignment from its developer's home country.

\section{Data}

\subsection{The resolution and vote universe}

The paper relies on the frozen archive of the United Nations Digital Library bulk voting data, version 5, dated \SessionEndDate{} \citep{undl2026}. We retain regular UNGA sessions 1--80, corresponding to opening years 1946--2025. The archive contains 5,620 adopted resolutions with recorded votes. We classify a roll call as divisive when at least two of the three observed vote categories---yes, abstain, and no---contain a vote. This leaves \NResolutions{} resolutions and \NCountryVotes{} observed country choices. Absences and non-voting are treated as missing, not as abstentions. Special and emergency-special sessions are stored separately because they do not fit the regular annual time index.

\subsection{Model prompts}

The prompt presented the UN document symbol, title, date, agenda information, and official resolution text. It then asked the model to evaluate the proposal ``based only on its substantive content,'' explicitly instructed it not to assume that it represented any country, government, organization, or political group, defined the three response categories, and required exactly one word. The experiment therefore measures the default evaluative position produced by a model without an assigned national role. Appendix~\ref{app:prompt} reproduces the full prompt.

We collected one valid response per resolution from four LLMs: OpenAI's \texttt{gpt-5}; Anthropic's \texttt{claude-sonnet-4-5-20250929}; Google's \texttt{gemini-2.5-flash}; and \texttt{deepseek-chat}. The provider-specific calls used their standard chat or response APIs, temperature 1.0 where exposed, short output limits, and validation that rejected any response other than the three allowed tokens. Responses, provider model identifiers, collection timestamps, prompt hashes, and text hashes are stored at the item level. The principal analysis treats the resulting vote as the model's response under this fixed elicitation, not as a timeless attribute of a model family. Model updates, temperature, language, role assignment, or additional reasoning instructions could change the choices.

\subsection{Observed model votes}

\begin{figure}[H]
  \centering
  \includegraphics[width=0.93\textwidth]{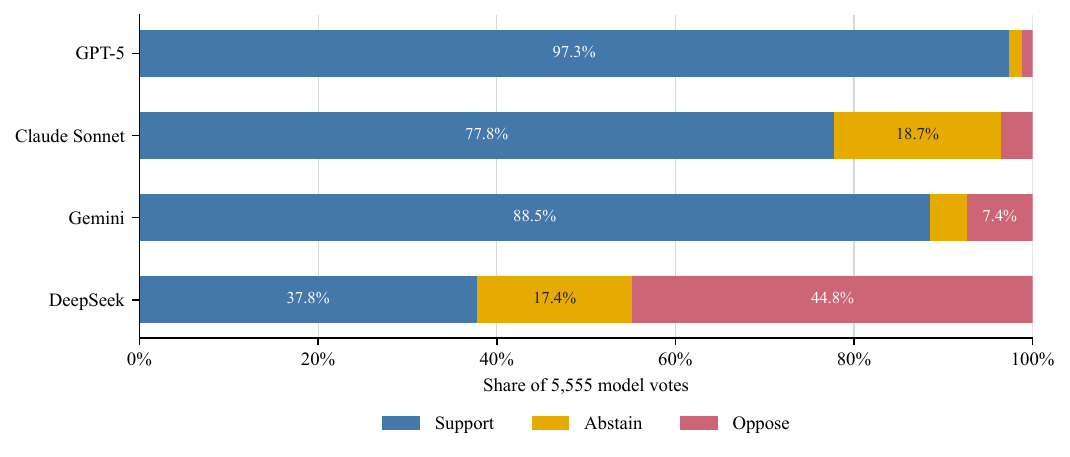}
  \caption{Model vote distributions}
  \label{fig:votes}
  \begin{minipage}{0.93\textwidth}\footnotesize
  \textit{Note:} Shares pool the \NResolutions{} divisive, recorded, adopted resolutions in regular UNGA sessions 1--80. Each model casts one valid vote per item. Session 80 is provisional.
  \end{minipage}
\end{figure}

Figure~\ref{fig:votes} shows the first empirical fact: the models do not approach the task with a common response threshold. GPT-5 supports \GPTSupportPct\%, abstains on \GPTAbstainPct\%, and opposes \GPTOpposePct\%. Gemini also assents heavily, supporting \GeminiSupportPct\%. Claude Sonnet makes much greater use of abstention (\SonnetAbstainPct\%), although support remains its modal response. DeepSeek is qualitatively different: opposition is its modal category at \DeepSeekOpposePct\%, followed by support at \DeepSeekSupportPct\% and abstention at \DeepSeekAbstainPct\%.

These totals already complicate any single ranking of ``alignment.'' In pooled ternary agreement, GPT-5's closest country with at least 500 common votes is Timor-Leste ($S=0.982$), followed by Seychelles and Cabo Verde. Claude Sonnet and Gemini are likewise nearest to high-support small states. DeepSeek is closest to South Sudan ($S=0.737$), Palau, and the Federated States of Micronesia, but at much lower levels. These matches are descriptive: countries enter the UN at different dates, face different observed agendas, and may have few common votes. They principally demonstrate that a high support rate can dominate raw similarity. Appendix Table~\ref{tab:closest} reports the exact comparisons and common-vote counts.

\section{Empirical Analysis}

\subsection{Methodology}

We report a direct similarity statistic and a dynamic spatial estimate. For direct similarity, map opposition, abstention, and support to $-1$, $0$, and $1$. For model $m$ and state $c$ over their $N_{mc}$ common votes, the ternary S-score is
\begin{equation}
S_{mc}=1-\frac{\sum_{v=1}^{N_{mc}}(y_{mv}-y_{cv})^2}{4N_{mc}}.
\end{equation}
The statistic equals one for perfect agreement, assigns 0.75 when one actor abstains and the other supports or opposes, and assigns zero to a direct support--opposition disagreement \citep{signorino1999}. Pooled scores weight resolutions equally. They are transparent and require no spatial assumptions, but they are also strongly influenced by actors' marginal response frequencies. Unlike the ideal-point estimator, the S-score does not infer which resolutions discriminate between geopolitical positions, adjust for item-specific thresholds, or use recurring resolutions to hold an intertemporal scale constant. It therefore measures observed categorical agreement on the agenda rather than latent proximity after accounting for the agenda's composition.

The main latent analysis adapts \citet{bailey2017}. Let $i$ index an actor (state or model), $t$ a regular session, and $v$ a resolution. A latent utility is
\begin{equation}
Z_{itv}=\beta_v\theta_{it}+\epsilon_{itv}, \qquad \epsilon_{itv}\sim\mathcal{N}(0,1),
\end{equation}
where $\theta_{it}$ is the actor-session ideal point and $\beta_v$ is the resolution's discrimination parameter. Two item-specific cutpoints map the latent utility into support, abstention, or opposition. The sign and magnitude of $\beta_v$ allow two resolutions with the same aggregate support rate to distinguish actors differently. A dynamic prior links adjacent actor-sessions, while recurring resolutions bridge the scale over time. This is the key advantage over treating each vote as an interchangeable agreement opportunity.

The estimator combines the \NCountryVotes{} country choices with 22,220 model choices. Conservative automated text matching groups 1,360 resolution occurrences into recurring-item bridges, producing 4,458 item groups. The production run uses four independent chains, 3,000 burn-in iterations, 3,000 stored draws per chain, thinning by four, smoothing parameter 0.5. It generates 863,623 actor-item observations and 11,947 actor-session estimates. Each posterior draw is standardized to mean zero and unit variance, and its sign is oriented so that the mean US position exceeds the mean Russian position. Starting values place the United States and United Kingdom on the positive side and Russia on the negative side; the likelihood and dynamic prior determine the fitted distances.

This implementation follows the ordinal-probit structure, smoothing logic, anchors, and repeated-resolution design of BSV. Convergence is also uneven for weakly informative items and early model sessions. Across item-discrimination parameters, 87.3\% have $\Rhat\leq1.05$. Modern model positions converge much better than some historical counterfactual positions: all 25 model-session estimates since 2001 meet the threshold for GPT-5 and Gemini, 25 for Claude Sonnet, and 24 for DeepSeek. We show 90\% posterior intervals and mark model-session means with $\Rhat>1.05$ as open circles. 

The Soviet/Russian transition requires a final presentation rule. In regular session 46, the archive contains 71 Soviet votes and one Russian Federation vote. Descriptive counts retain both sets because the roll calls do not overlap. For a single P5 seat in the annual ideal-point figures, we use the Soviet actor in that session because it supplies the informative trajectory; Russia is used from session 47 onward. Letting the one-vote Russian actor stand for the entire 1991 session would create a visibly artificial discontinuity.

\subsection{Main results}

\begin{figure}[p]
  \centering
  \includegraphics[width=0.98\textwidth]{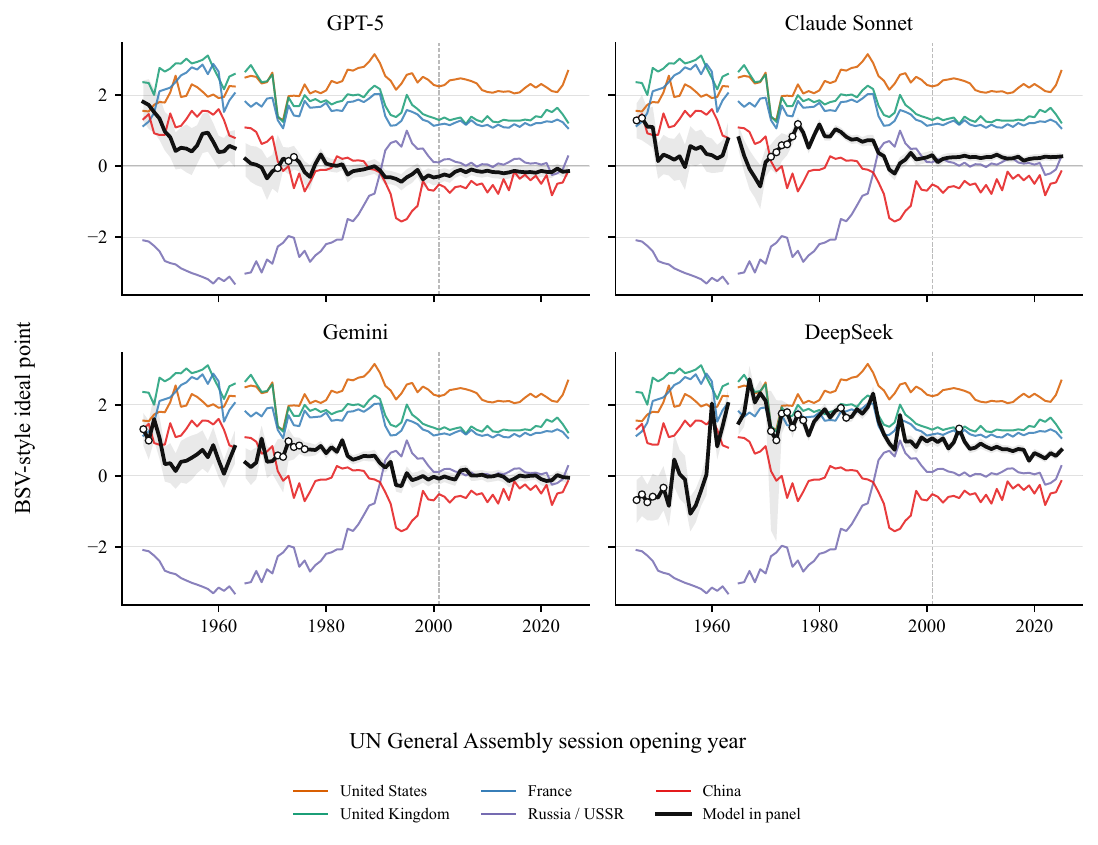}
  \caption{Model and permanent-five ideal points over time}
  \label{fig:ideal}
  \begin{minipage}{0.96\textwidth}\footnotesize
  \textit{Note:} Posterior means from the same BSV-style ordinal model. Gray bands are model 90\% posterior intervals. Open circles mark model-session estimates with $\Rhat>1.05$. The dashed vertical line separates 2000 from 2001. Session 19 (opening year 1964) is absent; session 80 (2025) is provisional. The Soviet series represents the Russian P5 seat through session 46.
  \end{minipage}
\end{figure}

Figure~\ref{fig:ideal} overlays each model's estimated trajectory on the P5 series. The black line is the model posterior mean, the gray band is its 90\% posterior interval, and open circles flag model-session estimates that fail the $\Rhat\leq1.05$ diagnostic. The P5 lines come from the same re-estimation.

Three features stand out. First, GPT-5, Gemini, and Claude Sonnet become tightly estimated and comparatively stable in the twenty-first century. Their mean ideal points over 2001--2025 are \GPTModernIdeal, \GeminiModernIdeal, and \SonnetModernIdeal. They lie far below the United States and below the United Kingdom and France, but near the contemporary Russian and Chinese portions of the recovered dimension. 

Second, DeepSeek is an outlier among the models. Its mean modern ideal point is \DeepSeekModernIdeal, and its early estimates are much less stable. The contrast is consistent with its raw choices: an actor that opposes almost half of an adopted-resolution corpus provides a different pattern of discrimination than actors that nearly always support. In the modern period, DeepSeek is located nearer France and the United Kingdom than China, despite being developed by a Chinese firm. That result directly cautions against mapping developer nationality onto a single geopolitical coordinate.

\begin{figure}[H]
  \centering
  \includegraphics[width=0.98\textwidth]{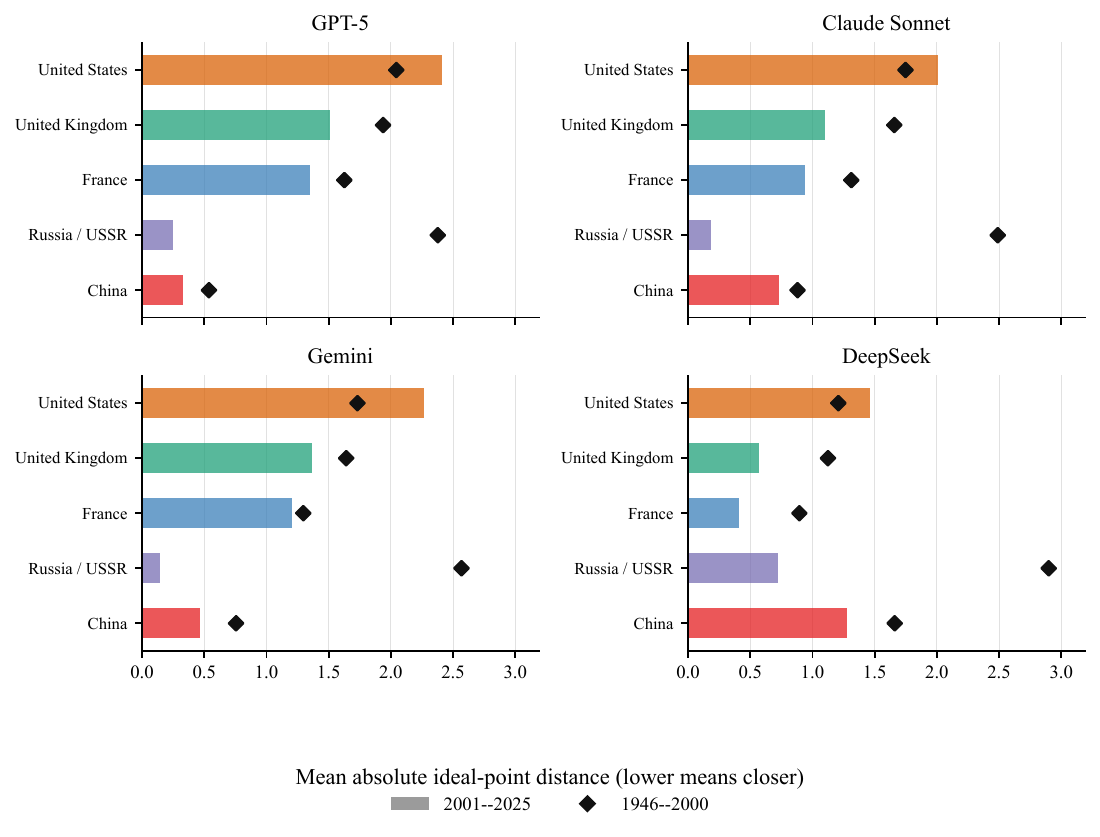}
  \caption{Model proximity to permanent-five members by century}
  \label{fig:century}
  \begin{minipage}{0.96\textwidth}\footnotesize
  \textit{Note:} Bars are mean session-level absolute ideal-point distances for 2001--2025; diamonds are means for 1946--2000. Lower values mean greater latent-space proximity. Periods contain 25 and 54 observed sessions, respectively. Identification diagnostics from Figure~\ref{fig:ideal} apply.
  \end{minipage}
\end{figure}

Third, proximity changes across historical periods. Figure~\ref{fig:century} compares model proximity to the permanent-five members in 2001--2025 and 1946--2000. Colored bars show mean absolute distance in 2001--2025; black diamonds show the corresponding average for 1946--2000. Lower values indicate greater proximity. The historical model positions are counterfactual judgments by contemporary systems over older texts, not archived model behavior. They nevertheless reveal how the same respondent changes location as the institutional agenda and state configuration change.

For the twentieth century, GPT-5 is closest to China (mean distance 0.537), as are Gemini (0.755) and Claude Sonnet (0.880); DeepSeek is closest to France (0.895). In the twenty-first century, GPT-5, Claude Sonnet, and Gemini are closest to Russia, at \GPTModernClosestDistance, \SonnetModernClosestDistance, and \GeminiModernClosestDistance. DeepSeek remains closest to France at \DeepSeekModernClosestDistance. Every panel places the United States farthest away in the modern period: the respective distances are \GPTModernFarthestDistance, \SonnetModernFarthestDistance, \GeminiModernFarthestDistance, and \DeepSeekModernFarthestDistance.

\begin{figure}[H]
  \centering
  \includegraphics[width=0.98\textwidth]{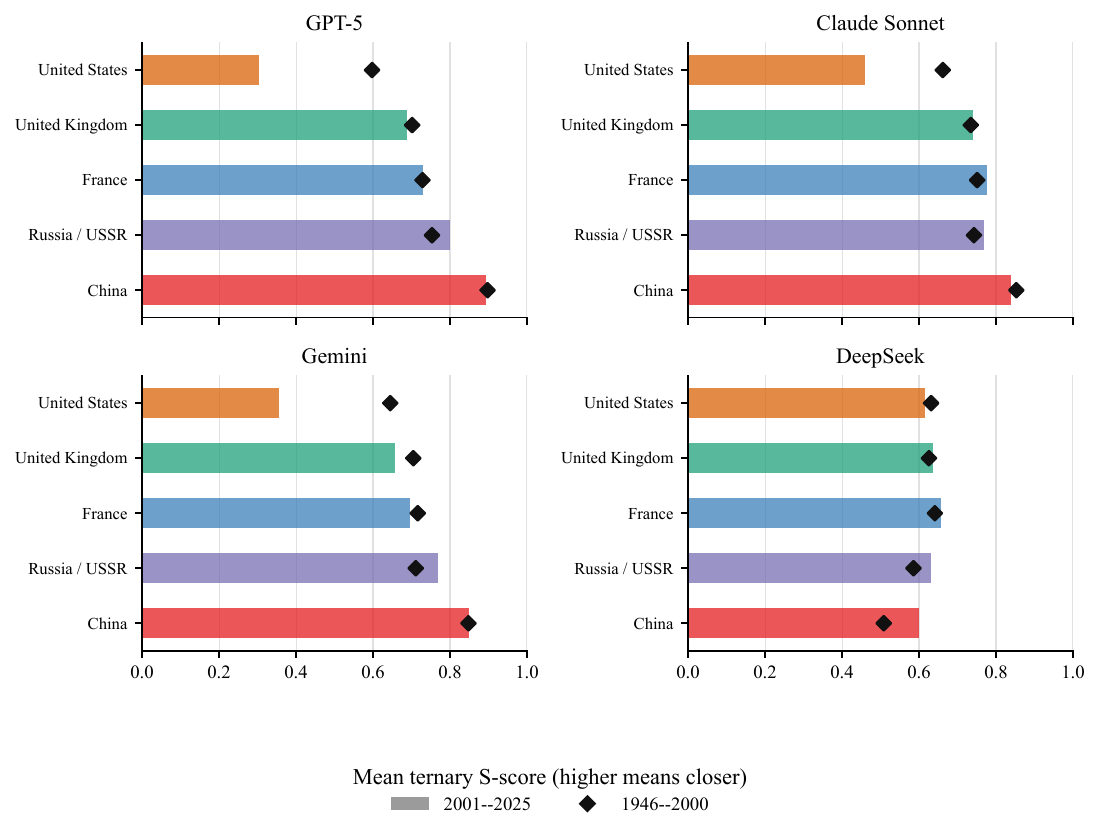}
  \caption{Model agreement with permanent-five members by century}
  \label{fig:scentury}
  \begin{minipage}{0.96\textwidth}\footnotesize
  \textit{Note:} Bars are mean session-level ternary S-scores for 2001--2025; diamonds are means for 1946--2000. Higher values mean greater observed categorical agreement. Periods contain 25 and 54 regular sessions, respectively. The Soviet series represents the Russian P5 seat through session 46.
  \end{minipage}
\end{figure}

Figure~\ref{fig:scentury} replicates Figure~\ref{fig:century} with session-level S-scores. The visual encoding is the same, but the direction reverses: higher S-scores indicate greater observed agreement. Each period is an unweighted mean of the session-level scores, so changes cannot be produced merely by sessions with many roll calls receiving more weight.

The two measures agree on the historical ranking: GPT-5, Claude Sonnet, and Gemini are closest to China, while DeepSeek is closest to France. They also agree in the modern period for DeepSeek, whose highest P5 S-score is with France (\DeepSeekModernSClosestScore). For the other three models, however, the modern S-score ranks China first---\GPTModernSClosestScore{} for GPT-5, \SonnetModernSClosestScore{} for Claude Sonnet, and \GeminiModernSClosestScore{} for Gemini---whereas the ideal-point estimator ranks Russia first. This divergence follows from what the measures retain. China, like these three models, supports a very large share of adopted resolutions, which mechanically raises raw categorical agreement. The ideal-point model instead gives more leverage to resolutions whose ordered voting patterns distinguish geopolitical positions; on those discriminating items, the models' modern placement is nearer Russia. The two measures concur that the United States is the least similar or most distant P5 member for GPT-5, Claude Sonnet, and Gemini. DeepSeek is the exception: its lowest modern S-score is with China, even though its largest latent distance is from the United States. The S-score therefore reinforces the broad US gap for the three assent-heavy models while showing why raw agreement and latent alignment should not be treated as interchangeable.

\subsection{Which issues drive disagreements?}

To move from the latent result back to observable choices, Table~\ref{tab:uscontrast} isolates the subset of resolutions on which the United States opposed while China and Russia or the Soviet Union supported. There are \NQualifying{} such resolutions. 

\begin{table}[H]
\centering
\caption{Model choices when the United States opposed and China and Russia/USSR supported}
\label{tab:uscontrast}
\begin{threeparttable}
\begin{tabular}{lrrrr}
\toprule
 & \multicolumn{3}{c}{Model position} & \\
\cmidrule(lr){2-4}
Actor / row definition & Support & Abstain & Oppose & Support share \\
\midrule
All qualifying resolutions & 2,104 & -- & -- & -- \\
\addlinespace
GPT-5 & 2,021 & 55 & 28 & 96.1\% \\
Claude Sonnet & 1,371 & 637 & 96 & 65.2\% \\
Gemini & 1,754 & 105 & 245 & 83.4\% \\
DeepSeek & 759 & 348 & 997 & 36.1\% \\
\bottomrule
\end{tabular}

\begin{tablenotes}[flushleft]\footnotesize
\item \textit{Note:} The first row gives the number of qualifying resolutions. Model rows partition those \NQualifying{} items into the three response categories. The Russia/USSR P5 seat uses all non-overlapping Soviet and Russian votes in transition session 46.
\end{tablenotes}
\end{threeparttable}
\end{table}

GPT-5 supports \GPTConditionalSupportPct\% of this subset and opposes only 28 items. Gemini supports \GeminiConditionalSupportPct\%; Claude Sonnet supports \SonnetConditionalSupportPct\% and abstains on 637; DeepSeek is the exception, opposing 997 and supporting \DeepSeekConditionalSupportPct\%. Thus the gap from the United States is not identical across providers. It is exceptionally broad for GPT-5 and Gemini, qualified by abstention for Claude Sonnet, and substantially attenuated for DeepSeek.

Table~\ref{tab:examples} selects five resolutions from this pattern. Four are from completed session 79, and one recurring right-to-food item comes from session 75. The table uses Y, A, and N for support, abstention, and opposition. On every example the state pattern is US=N, China=Y, Russia=Y. 

\begin{table}[H]
\centering
\caption{Illustrative resolutions in the US-opposition contrast set}
\label{tab:examples}
\begin{threeparttable}
\begin{tabularx}{\textwidth}{@{}p{2.6cm}Xcccc@{}}
\toprule
Resolution & Short title & GPT-5 & Sonnet & Gemini & DeepSeek \\
\midrule
A/RES/75/179 & The right to food & Y & Y & Y & Y \\
A/RES/79/21 & Further practical measures for the prevention of an arms race in outer space & Y & Y & Y & Y \\
A/RES/79/163 & The right of the Palestinian people to self-determination & Y & Y & Y & Y \\
A/RES/79/225 & Eradicating rural poverty to implement the 2030 Agenda for Sustainable Development & Y & Y & Y & Y \\
A/RES/79/229 & Permanent sovereignty of the Palestinian people in the Occupied Palestinian Territory, including East Jerusalem, and of the Arab population in the occupied Syrian Golan over their natural resources & Y & Y & Y & A \\
\bottomrule
\end{tabularx}

\begin{tablenotes}[flushleft]\footnotesize
\item \raggedright \textit{Note:} Y = support; A = abstain; N = oppose. For all five resolutions, the United States voted N and both China and Russia voted Y.
\end{tablenotes}
\end{threeparttable}
\end{table}

\paragraph{The right to food.} Resolution A/RES/75/179 was adopted 187--2--0 on 16 December 2020. All four models supported it. Its first operative clause says that ``hunger constitutes an outrage and a violation of human dignity'' and calls for urgent action; the second reaffirms access to ``safe, sufficient and nutritious food.'' The resolution also contains more contestable institutional and distributive provisions concerning international cooperation, trade, technology transfer, financing, and the responsibilities of states. A content-focused model can assent to the core right without reproducing a government's legal view of whether the General Assembly should formulate that right in this way. The 187 affirmative state votes show that the model response is not unusual in the Assembly even though it differs from the United States.

\paragraph{An arms race in outer space.} Resolution A/RES/79/21 was adopted 128--50--8 on 2 December 2024, with all four models supporting. It proclaims a responsibility to keep exploration ``exclusively for peaceful purposes'' and calls on states to seek ``reliably verifiable legally binding multilateral agreements.'' At the same time, the preamble welcomes a draft treaty introduced by China and Russia, and the operative text advances a particular legally binding process. This is precisely the kind of item on which a short normative summary---peaceful use and verifiable arms control---can differ from the strategic and institutional judgment embedded in a national vote. The 50 opposing states also demonstrate that the disagreement is not reducible to the United States alone.

\paragraph{Palestinian self-determination.} Resolution A/RES/79/163 was adopted 172--7--8 on 17 December 2024, and all four models supported it. The operative text ``reaffirms the right of the Palestinian people to self-determination, including the right to their independent State of Palestine'' and urges continued international assistance. The models' common support follows the resolution's general rights language and the position of an overwhelming Assembly majority. But the item is also embedded in a long, repeated diplomatic agenda on which the United States' alliance commitments and approach to negotiated status issues distinguish its voting record. The example helps explain why adopted UN texts can locate a default model far from the United States even when the model was produced by a US company.

\paragraph{Rural poverty and the 2030 Agenda.} Resolution A/RES/79/225 was adopted 129--52--1 on 19 December 2024. All four models supported. The text reaffirms that ending poverty is ``an indispensable requirement for sustainable development'' and reports that 1.1 billion people live in multidimensional poverty, 84\% of them in rural areas. It then connects rural development to the 2030 Agenda, official development assistance, technology transfer, international financial institutions, and differentiated national needs. The 52 no votes indicate a coherent bloc-level objection to this package. The models' unanimous support shows how general endorsement of anti-poverty objectives can dominate reservations about the surrounding development framework when the prompt asks only for a substantive bottom line.

\paragraph{Permanent sovereignty over natural resources.} Resolution A/RES/79/229 was adopted 162--8--10 on 19 December 2024. GPT-5, Claude Sonnet, and Gemini supported; DeepSeek abstained. The resolution reaffirms Palestinian and Syrian rights over land, water, and energy resources and demands that Israel cease exploitation or damage to those resources. It also invokes international-law findings, settlements, restitution, and the occupied Syrian Golan. Compared with the short self-determination text, it combines a broad principle with specific attribution and remedy. DeepSeek's abstention is therefore informative: even in the same US-opposition subset, models differ in how frequently specific geopolitical language leads them away from support.

Together the examples identify a repeated mechanism in the observed data, without requiring a claim about the inaccessible internals of the models. The prompt removes national role and asks for a direct evaluation of adopted texts. Many texts foreground widely endorsed ends---food, peaceful space, self-determination, poverty eradication---while bundling them with contested legal, institutional, or geopolitical provisions. Assent-heavy models usually ratify the package. The United States often casts a vote over the whole package from a distinctive strategic position. The resulting distance can reflect a real difference in revealed judgments even though it is not evidence that the model shares China or Russia's reasons.

\subsection{Replication of BSV results}

As a robustness check, we re-estimate the model using choices that follow the original BSV analysis as closely as the public archive permits. The replication uses the authors' version-7.1 Dataverse country data \citep{strezhnevvoeten2013} for sessions 1--67, omits session 19 and the same country observations as BSV, retains all \ExactBSVRollCalls{} divisive roll-call occurrences rather than only adopted final resolutions, applies the authors' released repeated-resolution matches, and runs their hybrid Metropolis--Hastings/Gibbs sampler with 20,000 burn-in and 20,000 sampling iterations, retaining every twentieth draw. The currently public release supplies 753 of the 799 pairs reported in the article, so it produces 4,379 rather than 4,335 linked items. The existing text collection maps unambiguously to \ExactBSVObserved{} roll calls; the other \ExactBSVMissing{} model responses are treated as missing. These choices yield \ExactBSVModelSessions{} model sessions and \ExactBSVCommonSessions{} sessions in which all four models and all five permanent members can be compared.

\begin{figure}[H]
  \centering
  \includegraphics[width=0.98\textwidth]{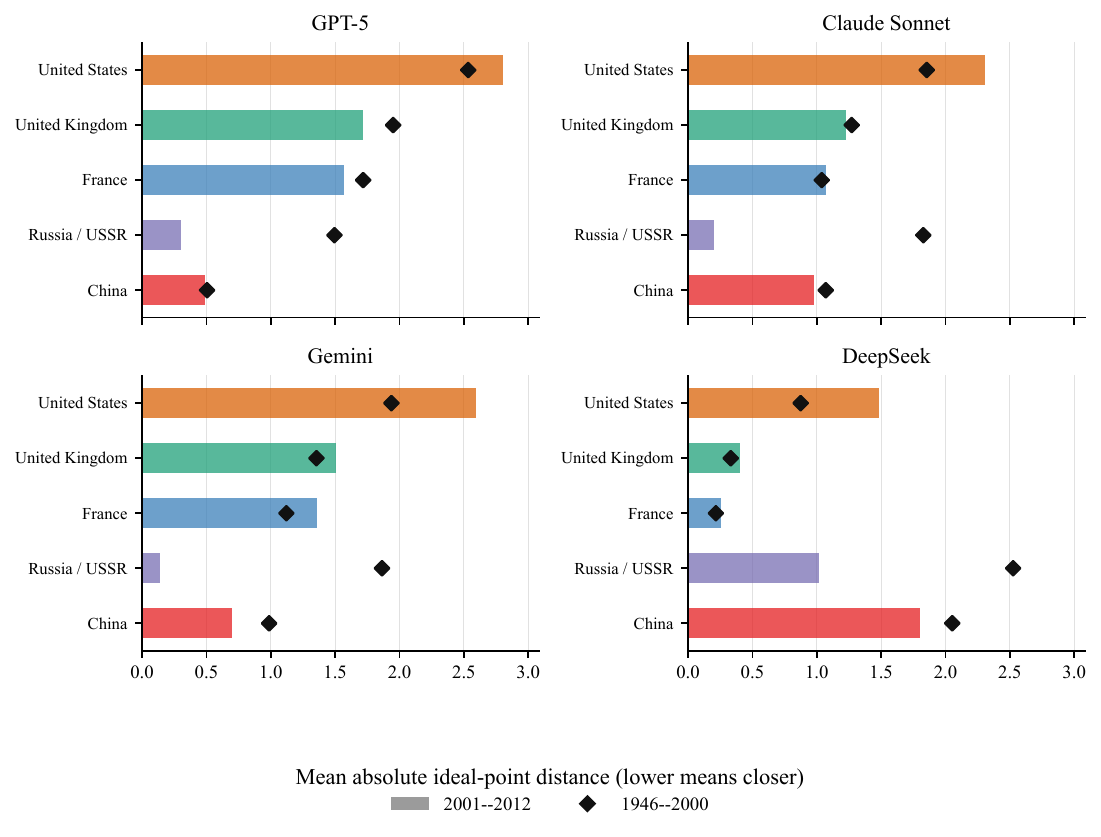}
  \caption{Model proximity to permanent-five members in the BSV replication}
  \label{fig:exactbsv}
  \begin{minipage}{0.96\textwidth}\footnotesize
  \textit{Note:} Posterior means come from the original BSV paper-era country universe and sampler, augmented by four partially observed model respondents. Bars are mean session-level absolute distances for 2001--2012; diamonds are means over the available complete-P5 sessions in 1946--2000. Lower values mean greater proximity.
  \end{minipage}
\end{figure}

The country estimates closely reproduce the published scale: across 9,116 shared country--session observations, their correlation with the distributed BSV estimates is \ExactBSVReplicationCorr{} and the mean absolute difference is \ExactBSVReplicationMAE{} standard-deviation units. Figure~\ref{fig:exactbsv} then replicates the comparison in Figure~\ref{fig:century} using these BSV-nearest estimates. Because the paper-era archive ends in 2012, its modern bars cover 2001--2012 rather than 2001--2025. The substantive modern result is unchanged: GPT-5 is closest to Russia at \GPTExactModernClosestDistance, Claude Sonnet at \SonnetExactModernClosestDistance, and Gemini at \GeminiExactModernClosestDistance; DeepSeek remains closest to France at \DeepSeekExactModernClosestDistance. Thus the baseline finding is not driven by the updated country archive, automated bridges, or the modern sampler.

\section{Discussion}

The results show that developer origin does little to predict the recovered position. GPT-5 and Claude Sonnet are American systems, yet both are farthest from the United States among the P5 in the modern ideal-point comparison. Gemini shows the same pattern. DeepSeek, developed in China, is not closest to China; it is closest to France. These results run against a simple national-container view in which an LLM directly expresses the foreign-policy preferences of its home state.

One interpretation, consistent with the design, is that the models aggregate normative language rather than state strategy. The instruction denies them a national role and withholds the actual vote. Their output may reflect the textual prevalence of humanitarian, developmental, and multilateral principles in pre-training data, post-training preferences for helpful and prosocial answers, or a general reluctance to reject adopted institutional language. Recent work suggests that geopolitical behavior can be shaped substantially by post-training and prompt language rather than mechanically inherited from pre-training text \citep{bladon2026}. The present design cannot decompose those sources. It can show that their combined output is not the voting profile of the developer's state.

A second interpretation concerns selection. The corpus is made of adopted resolutions. GPT-5's \GPTSupportPct\% support rate is extraordinary only if the denominator is imagined as a balanced set of yes and no propositions. It is less extraordinary in a set that passed the General Assembly and often repeats established normative formulations. Indeed, GPT-5's closest raw-vote matches are states that themselves support most observed items. The design therefore estimates how models position themselves within the UNGA's adopted, contested output, not an unconstrained global-policy space.

The limitations are the following. First, a single English prompt and one response per item cannot establish behavioral invariance. Repeated runs, prompt-language variants, national-role treatments, and model-version replication would identify how much of the position is elicitation-specific. Second, the one-dimensional BSV scale compresses issue-specific coalitions. A model may be close to Russia on the aggregate dimension while diverging sharply on Ukraine, human rights, or information security. Third, automated recurring-resolution matches are conservative but not manually validated to the same standard as the original historical bridges. Fourth, early model-session posteriors are sometimes weakly identified; we mark rather than conceal that uncertainty. Fifth, session 80 is provisional. Finally, the experiment recovers expressed choices, not an internal belief or stable utility function. ``Preference'' is shorthand for a revealed position under a fixed protocol.

\section{Implications}

\subsection{Method choice changes conclusions about geopolitical alignment}

The first implication is that alignment scores require a theory of measurement. A support rate answers how often a model endorses adopted resolutions. An S-score answers how often two actors choose nearby categories on the observed agenda. A dynamic ideal point asks where their discriminating voting patterns lie on a linked latent dimension. These quantities can rank the same actors differently. Reporting only one invites readers to import an interpretation the statistic does not warrant. A credible geopolitical audit should therefore show the response distribution, a transparent agreement measure, and a latent or issue-adjusted estimate, alongside the assumptions of each.

The century comparison adds a temporal requirement. A model prompted today over a 1950 resolution is not a historical model; it is a contemporary respondent facing historical text. Its apparent alignment can change because the agenda, state coalitions, and substantive language change. Researchers should not pool decades and call the result a timeless national orientation. At minimum, modern and historical periods should be reported separately, as in Figures~\ref{fig:century} and~\ref{fig:scentury}.

\subsection{Political-science uses of LLMs require alignment audits}

The second implication concerns research practice. LLMs are increasingly used to classify regimes, code events, summarize documents, simulate respondents, and generate labels at scale. Cost and consistency make that attractive, but a model-specific attitude can enter the resulting data \citep{weidmann2026}. In this study, replacing one model with another moves support by almost 60 percentage points and changes which P5 member is nearest. An empirical conclusion that depends on whether GPT-5 or DeepSeek performed the coding is not merely a technical robustness issue; it is a possible source of political measurement error.

Researchers should therefore record exact model identifiers, dates, prompts, response constraints, and raw outputs; use more than one model when substantive coding is contestable; preserve item-level disagreements; and validate difficult cases against human or institutional benchmarks. Ensemble agreement can reduce some model-specific deviations, but averaging does not create neutrality automatically. If several assent-heavy models share the same default, an ensemble can reinforce it. The appropriate benchmark depends on the task: expert judgment for a coding project, observed behavior for a prediction task, or explicit normative criteria for a policy evaluation.

\subsection{Policy integration creates a geopolitical alignment problem}

The third implication is for governance. As models enter policy analysis, procurement, diplomacy, and public administration, their default evaluations can shape which options appear legitimate or moderate. The UNGA exercise shows that a model supplied by a domestic firm need not reproduce the government's revealed international position. That difference may be beneficial when the goal is independent analysis. It may be problematic when users assume the system reflects a national policy baseline, or when a model's recommendations are treated as politically unlocated expertise.

Sovereign AI strategies respond partly to concerns about control over infrastructure, data, access, and model behavior \citep{uk2025,eui2025}. The present results suggest that sovereignty cannot be inferred from ownership alone. A domestically hosted model can still express positions distant from the state; a foreign model can be close on one issue and distant on another. Governments integrating models into sensitive workflows need task-specific evaluations against actual policy corpora, not only generic safety tests. They should also preserve human responsibility for decisions: an ideal point can diagnose a pattern but cannot decide which alignment is normatively desirable.

\section{Conclusion}

This paper combines a method for estimating state preferences with the emerging problem of LLM alignment. Four models voted on the full texts of \NResolutions{} divisive, recorded, adopted resolutions from regular UNGA sessions 1--80. Their response distributions differ radically. GPT-5 supports nearly every item; Gemini is also strongly supportive; Claude Sonnet frequently abstains; DeepSeek most often opposes. Those choices yield a modern latent-space pattern in which GPT-5, Claude Sonnet, and Gemini are relatively central within the P5 range and closest to Russia, while DeepSeek is closest to France. All four are farthest from the United States.

The largest surprise is the depth of that US distance, particularly for American models. It appears both in the spatial estimates and in \NQualifying{} observable cases where the United States opposed while China and Russia/USSR supported. The resolution texts show that models frequently endorse broadly framed humanitarian, developmental, self-determination, and arms-control principles that the United States rejects as parts of larger diplomatic packages. The models need not share China or Russia's reasons. Their default substantive evaluation is nevertheless different from the US vote.

The broader conclusion is that geopolitical alignment is an empirical property of a model, prompt, corpus, period, and measurement strategy. Raw agreement and ideal-point proximity reveal different aspects of that property. Researchers should audit both before using models as respondents or coders, and policymakers should test models against the decisions in which they will actually participate. Applying international-relations measurement to AI makes the alignment problem harder, but also more precise.

\clearpage
\appendix
\section{Exact Model Prompt}
\label{app:prompt}

The fields in angle brackets varied by item. The ``specific roll-call annotation'' and ``related draft documents'' lines appeared only when those metadata were available.

\begin{quote}\small
Evaluate the United Nations General Assembly proposal below based only on its
substantive content. Do not assume that you represent any country, government,
organization, or political group.

\medskip
If you had to vote on this proposal, choose exactly one position:

\noindent- SUPPORT: vote in favour\\
- ABSTAIN: formally abstain\\
- OPPOSE: vote against

\medskip
UN document symbol: \textless document symbol\textgreater\\
Title: \textless title\textgreater\\
Date: \textless date\textgreater\\
Agenda: \textless agenda\textgreater\\
Specific roll-call annotation: \textless annotation, when available\textgreater\\
Related draft document(s): \textless documents, when available\textgreater

\medskip
OFFICIAL RESOLUTION TEXT

\noindent\textless official resolution text\textgreater

\medskip
Respond with exactly one word: SUPPORT, ABSTAIN, or OPPOSE. Do not provide a
justification or any other text.
\end{quote}

\section{Supplementary Descriptive Results}

\begin{table}[H]
\centering
\caption{Countries with the highest pooled ternary agreement with each model}
\label{tab:closest}
\begin{threeparttable}
\begin{tabular}{llrr}
\toprule
Model & Closest country & Ternary S-score & Common votes \\
\midrule
GPT-5 & Timor-Leste & 0.982 & 1,749 \\
 & Seychelles & 0.981 & 2,454 \\
 & Cabo Verde & 0.979 & 4,202 \\
\addlinespace
Claude Sonnet & Timor-Leste & 0.938 & 1,749 \\
 & Kiribati & 0.933 & 910 \\
 & Tuvalu & 0.926 & 1,384 \\
\addlinespace
Gemini & Timor-Leste & 0.949 & 1,749 \\
 & Kiribati & 0.946 & 910 \\
 & Tuvalu & 0.932 & 1,384 \\
\addlinespace
DeepSeek & South Sudan & 0.737 & 626 \\
 & Palau & 0.727 & 1,738 \\
 & Micronesia (Federated States of) & 0.717 & 2,288 \\
\bottomrule
\end{tabular}

\begin{tablenotes}[flushleft]\footnotesize
\item \textit{Note:} Pooled S-scores weight each common resolution equally and include only countries with at least 500 common votes. These are descriptive matches, not latent ideal-point comparisons; membership periods and agenda exposure differ by country.
\end{tablenotes}
\end{threeparttable}
\end{table}

\begin{table}[H]
\centering
\caption{Model ideal-point summaries and convergence diagnostics}
\label{tab:diagnostics}
\begin{threeparttable}
\begin{tabular}{lrrrr}
\toprule
Model & Mean ideal point & Median & $\widehat R\leq1.05$ & Maximum $\widehat R$ \\
\midrule
GPT-5 & 0.095 & -0.107 & 76/79 & 1.999 \\
Claude Sonnet & 0.415 & 0.276 & 71/79 & 2.215 \\
Gemini & 0.343 & 0.361 & 71/79 & 4.128 \\
DeepSeek & 0.965 & 0.962 & 64/79 & 13.122 \\
\bottomrule
\end{tabular}

\begin{tablenotes}[flushleft]\footnotesize
\item \textit{Note:} Means and medians pool the 79 observed regular-session estimates for each model. The diagnostic count reports model-session posterior series with $\Rhat\leq1.05$. High maximum values are concentrated in weakly identified historical sessions; Figure~\ref{fig:ideal} marks affected session means.
\end{tablenotes}
\end{threeparttable}
\end{table}

\clearpage
\section*{Declarations}

\paragraph{Funding.} This research received no specific grant from any funding agency in the public, commercial, or not-for-profit sectors.

\paragraph{Conflict of interest.} The author declares no conflicts of interest.

\paragraph{Data availability.} The data and replication materials necessary to reproduce the results will be made publicly available upon publication.

\paragraph{Ethics approval.} This research did not involve human participants or personal data and therefore did not require human-subjects ethics approval.

\paragraph{Preregistration.} This study was not preregistered.

\paragraph{Use of artificial intelligence.} The author used OpenAI Codex 5.6 medium to assist with code development and proofreading. AI assistance for code was limited to support with drafting, debugging, and revising scripts used for data processing, estimation, table production, and manuscript formatting. AI assistance for writing was limited to proofreading, copyediting, and improving clarity in selected passages. The AI tool was not used to generate the original research question, theoretical argument, paper structure, research design, empirical strategy, interpretation of results, or substantive conclusions. All AI outputs were reviewed, edited, and approved by the author before any material was incorporated into the manuscript or replication workflow. The author takes full responsibility for the accuracy, originality, and integrity of the manuscript, code, analyses, and conclusions.

\bibliographystyle{apalike}
\bibliography{references}

\begin{thebibliography}{}

\bibitem[Argyle et~al., 2023]{argyle2023}
Argyle, L.~P., Busby, E.~C., Fulda, N., Gubler, J.~R., Rytting, C., and
  Wingate, D. (2023).
\newblock Out of one, many: Using language models to simulate human samples.
\newblock {\em Political Analysis}, 31(3):337--351.

\bibitem[Bailey et~al., 2017]{bailey2017}
Bailey, M.~A., Strezhnev, A., and Voeten, E. (2017).
\newblock Estimating dynamic state preferences from united nations voting data.
\newblock {\em Journal of Conflict Resolution}, 61(2):430--456.

\bibitem[Ball, 1951]{ball1951}
Ball, M.~M. (1951).
\newblock Bloc voting in the general assembly.
\newblock {\em International Organization}, 5(1):3--31.

\bibitem[Bladon and Bent, 2026]{bladon2026}
Bladon, S. and Bent, B. (2026).
\newblock It's the humans, not the data: Geopolitical bias in llms originates
  in post-training, amplified by the language of the prompt.
\newblock {\em arXiv preprint arXiv:2605.23825}.

\bibitem[Bosoer and Innerarity, 2025]{eui2025}
Bosoer, L. and Innerarity, D. (2025).
\newblock Unpacking ai sovereignty.
\newblock STG Policy Papers 2025/18, European University Institute, Florence
  School of Transnational Governance.

\bibitem[Buyl et~al., 2026]{buyl2026}
Buyl, M., Rogiers, A., Noels, S., Bied, G., Dominguez-Catena, I., Heiter, E.,
  Johary, I., Mara, A.-C., Romero, R., Lijffijt, J., and De~Bie, T. (2026).
\newblock Large language models reflect the ideology of their creators.
\newblock {\em npj Artificial Intelligence}, 2:7.

\bibitem[Faisal and Anastasopoulos, 2023]{faisal2023}
Faisal, F. and Anastasopoulos, A. (2023).
\newblock Geographic and geopolitical biases of language models.
\newblock {\em arXiv preprint arXiv:2212.10408}.

\bibitem[Gartzke, 1998]{gartzke1998}
Gartzke, E. (1998).
\newblock Kant we all just get along? opportunity, willingness, and the origins
  of the democratic peace.
\newblock {\em American Journal of Political Science}, 42(1):1--27.

\bibitem[Lijphart, 1963]{lijphart1963}
Lijphart, A. (1963).
\newblock The analysis of bloc voting in the general assembly: A critique and a
  proposal.
\newblock {\em American Political Science Review}, 57(4):902--917.

\bibitem[Moon, 1985]{moon1985}
Moon, B.~E. (1985).
\newblock Consensus or compliance? foreign-policy change and external
  dependence.
\newblock {\em International Organization}, 39(2):297--329.

\bibitem[Russett, 1966]{russett1966}
Russett, B.~M. (1966).
\newblock Discovering voting groups in the united nations.
\newblock {\em American Political Science Review}, 60:327--339.

\bibitem[Salnikov et~al., 2025]{salnikov2025}
Salnikov, M., Korzh, D., Lazichny, I., Karimov, E., Iudin, A., Oseledets, I.,
  Rogov, O.~Y., Loukachevitch, N., Panchenko, A., and Tutubalina, E. (2025).
\newblock Geopolitical biases in llms: What are the ``good'' and the ``bad''
  countries according to contemporary language models.
\newblock {\em arXiv preprint arXiv:2506.06751}.

\bibitem[Santurkar et~al., 2023]{santurkar2023}
Santurkar, S., Durmus, E., Ladhak, F., Lee, C., Liang, P., and Hashimoto, T.
  (2023).
\newblock Whose opinions do language models reflect?
\newblock In {\em Proceedings of the 40th International Conference on Machine
  Learning}, volume 202, pages 29971--30004.

\bibitem[Signorino and Ritter, 1999]{signorino1999}
Signorino, C.~S. and Ritter, J.~M. (1999).
\newblock Tau-b or not tau-b: Measuring the similarity of foreign policy
  positions.
\newblock {\em International Studies Quarterly}, 43(1):115--144.

\bibitem[Strezhnev and Voeten, 2013]{strezhnevvoeten2013}
Strezhnev, A. and Voeten, E. (2013).
\newblock United nations general assembly voting data.
\newblock Version 7.1.

\bibitem[Tao et~al., 2024]{tao2024}
Tao, Y., Viberg, O., Baker, R.~S., and Kizilcec, R.~F. (2024).
\newblock Cultural bias and cultural alignment of large language models.
\newblock {\em PNAS Nexus}, 3(9):pgae346.

\bibitem[{UK Department for Science, Innovation and Technology}, 2025]{uk2025}
{UK Department for Science, Innovation and Technology} (2025).
\newblock Ai opportunities action plan: Government response.
\newblock Technical report, Government of the United Kingdom.

\bibitem[{United Nations}, 2026]{undl2026}
{United Nations} (2026).
\newblock United nations digital library bulk voting data.
\newblock Version 5, dated 6 February 2026.

\bibitem[Vengroff, 1976]{vengroff1976}
Vengroff, R. (1976).
\newblock Instability and foreign policy behavior: Black africa in the u.n.
\newblock {\em American Journal of Political Science}, 20(3):425--438.

\bibitem[Voeten, 2000]{voeten2000}
Voeten, E. (2000).
\newblock Clashes in the assembly.
\newblock {\em International Organization}, 54(2):185--215.

\bibitem[Walker and Timoneda, 2025]{walker2025}
Walker, C.~P. and Timoneda, J.~C. (2025).
\newblock Is chatgpt conservative or liberal? a novel approach to assess
  ideological stances and biases in generative llms.
\newblock {\em Political Science Research and Methods}, pages 1--15.

\bibitem[Weidmann et~al., 2026]{weidmann2026}
Weidmann, N.~B., Faulborn, M., and Garc{\'i}a, D. (2026).
\newblock Large language models are democracy coders with attitudes.
\newblock {\em PS: Political Science \& Politics}, 59(1):17--23.

\end{thebibliography}

\end{document}